\documentclass[12pt]{article}
\usepackage{authblk}
\usepackage[bookmarksnumbered, colorlinks, plainpages]{hyperref}
\usepackage{amsmath, amsthm, amscd, amsfonts, amssymb, graphicx, color, booktabs}
\usepackage{siunitx}
\DeclareSIUnit\barn{b}
\usepackage{ifthen} 
\newboolean{uprightparticles}
\setboolean{uprightparticles}{false} 
\usepackage{xspace} 
\usepackage{upgreek}

\def\lhcb   {\mbox{LHCb}\xspace}

\def\MagUp {\mbox{\em Mag\kern -0.05em Up}\xspace}

\ifthenelse{\boolean{uprightparticles}}%
{

 \def\Pmu         {\ensuremath{\upmu}\xspace}

 \def\Ppi         {\ensuremath{\uppi}\xspace}

 \def\PDelta      {\ensuremath{\Delta}\xspace}                 
 \def\PXi         {\ensuremath{\Xi}\xspace}                 
 \def\PLambda     {\ensuremath{\Lambda}\xspace}                 
 \def\PSigma      {\ensuremath{\Sigma}\xspace}                 
 \def\POmega      {\ensuremath{\Omega}\xspace}                 
 \def\PUpsilon    {\ensuremath{\Upsilon}\xspace}
 \let\oldPi\Pi
 \def\PPi         {\ensuremath{\oldPi}\xspace}

 \def\PB      {\ensuremath{\mathrm{B}}\xspace}                 
                  
 \def\PD      {\ensuremath{\mathrm{D}}\xspace}

 \def\PK      {\ensuremath{\mathrm{K}}\xspace}

 \def\Pc      {\ensuremath{\mathrm{c}}\xspace}

 \def\Ph      {\ensuremath{\mathrm{h}}\xspace}                 
 \def\Pi      {\ensuremath{\mathrm{i}}\xspace}

 \def\Pp      {\ensuremath{\mathrm{p}}\xspace}

 \def\Ps      {\ensuremath{\mathrm{s}}\xspace}

 \def\thebaroffset{0.0em}
}
{

 \def\Pmu         {\ensuremath{\mu}\xspace}

 \def\Ppi         {\ensuremath{\pi}\xspace}

 \mathchardef\PDelta="7101
 \mathchardef\PXi="7104
 \mathchardef\PLambda="7103
 \mathchardef\PSigma="7106
 \mathchardef\POmega="710A
 \mathchardef\PUpsilon="7107
 \mathchardef\PPi="7105
                  
 \def\PB      {\ensuremath{B}\xspace}                 
                  
 \def\PD      {\ensuremath{D}\xspace}

 \def\PK      {\ensuremath{K}\xspace}

 \def\Pc      {\ensuremath{c}\xspace}

 \def\Ph      {\ensuremath{h}\xspace}                 
 \def\Pi      {\ensuremath{i}\xspace}

 \def\Pp      {\ensuremath{p}\xspace}

 \def\Ps      {\ensuremath{s}\xspace}

 \def\thebaroffset{0.18em}
}
\newcommand{\offsetoverline}[2][\thebaroffset]{\kern #1\overline{\kern -#1 #2}}%

\makeatletter
\ifcase \@ptsize \relax
  \newcommand{\miniscule}{\@setfontsize\miniscule{4}{5}}
\or
  \newcommand{\miniscule}{\@setfontsize\miniscule{5}{6}}
\or
  \newcommand{\miniscule}{\@setfontsize\miniscule{5}{6}}
\fi
\makeatother

\DeclareRobustCommand{\optbar}[1]{\shortstack{{\miniscule (\rule[.5ex]{1.25em}{.18mm})}
  \\ [-.7ex] $#1$}}

\def\mup        {{\ensuremath{\Pmu^+}}\xspace}
\def\mun        {{\ensuremath{\Pmu^-}}\xspace} 

\def\mumu       {{\ensuremath{\Pmu^+\Pmu^-}}\xspace}

\def\squark    {{\ensuremath{\Ps}}\xspace}

\def\cquark    {{\ensuremath{\Pc}}\xspace}

\def\pion   {{\ensuremath{\Ppi}}\xspace}

\def\pip    {{\ensuremath{\pion^+}}\xspace}
\def\pim    {{\ensuremath{\pion^-}}\xspace}

\def\kaon    {{\ensuremath{\PK}}\xspace}

\def\KorKbar {\kern \thebaroffset\optbar{\kern -\thebaroffset \PK}{}\xspace}

\def\Kp      {{\ensuremath{\kaon^+}}\xspace}
\def\Km      {{\ensuremath{\kaon^-}}\xspace}

\def\KS      {{\ensuremath{\kaon^0_{\mathrm{S}}}}\xspace}

\def\KL      {{\ensuremath{\kaon^0_{\mathrm{L}}}}\xspace}

\def\D       {{\ensuremath{\PD}}\xspace}

\def\DorDbar {\kern \thebaroffset\optbar{\kern -\thebaroffset \PD}\xspace}
\def\Dz      {{\ensuremath{\D^0}}\xspace}

\def\Dp      {{\ensuremath{\D^+}}\xspace}
\def\Dm      {{\ensuremath{\D^-}}\xspace}

\def\DpDm    {\ensuremath{\Dp {\kern -0.16em \Dm}}\xspace}

\def\Dstarz  {{\ensuremath{\D^{*0}}}\xspace}

\def\Dsp     {{\ensuremath{\D^+_\squark}}\xspace}

\def\B       {{\ensuremath{\PB}}\xspace}

\def\BorBbar {\kern \thebaroffset\optbar{\kern -\thebaroffset \PB}\xspace}

\def\Bd      {{\ensuremath{\B^0}}\xspace}

\def\BdorBdbar {\kern \thebaroffset\optbar{\kern -\thebaroffset \Bd}\xspace}

\def\Bs      {{\ensuremath{\B^0_\squark}}\xspace}

\def\BsorBsbar {\kern \thebaroffset\optbar{\kern -\thebaroffset \Bs}\xspace}

\def\Y#1S{\ensuremath{\PUpsilon{(#1S)}}\xspace}

\def\proton      {{\ensuremath{\Pp}}\xspace}

\def\Lz          {{\ensuremath{\PLambda}}\xspace}

\def\LorLbar     {\kern \thebaroffset\optbar{\kern -\thebaroffset \PLambda}\xspace}

\def\Lc          {{\ensuremath{\Lz^+_\cquark}}\xspace}

\def\BF         {{\ensuremath{\mathcal{B}}}\xspace}

\newcommand{\decay}[2]{\ensuremath{#1\!\to #2}\xspace} 

\def\to                 {\ensuremath{\rightarrow}\xspace}

\def\CP                {{\ensuremath{C\!P}}\xspace}

\def\AT#1     {\ensuremath{A_{\mathrm{T}}^{#1}}\xspace}           

\def\C#1      {\ensuremath{\mathcal{C}_{#1}}\xspace}                       
\def\Cp#1     {\ensuremath{\mathcal{C}_{#1}^{'}}\xspace}                    
\def\Ceff#1   {\ensuremath{\mathcal{C}_{#1}^{\mathrm{(eff)}}}\xspace}        
\def\Cpeff#1  {\ensuremath{\mathcal{C}_{#1}^{'\mathrm{(eff)}}}\xspace}       
\def\Ope#1    {\ensuremath{\mathcal{O}_{#1}}\xspace}                       
\def\Opep#1   {\ensuremath{\mathcal{O}_{#1}^{'}}\xspace}                    

\newcommand{\aunit}[1]{\ensuremath{\text{\,#1}}}       

\newcommand{\tev}{\aunit{Te\kern -0.1em V}\xspace}
\newcommand{\gev}{\aunit{Ge\kern -0.1em V}\xspace}
\newcommand{\mev}{\aunit{Me\kern -0.1em V}\xspace}
\newcommand{\kev}{\aunit{ke\kern -0.1em V}\xspace}
\newcommand{\ev}{\aunit{e\kern -0.1em V}\xspace}
 
\newcommand{\mevc}{\ensuremath{\aunit{Me\kern -0.1em V\!/}c}\xspace}
\newcommand{\gevc}{\ensuremath{\aunit{Ge\kern -0.1em V\!/}c}\xspace}
\newcommand{\mevcc}{\ensuremath{\aunit{Me\kern -0.1em V\!/}c^2}\xspace}
\newcommand{\gevcc}{\ensuremath{\aunit{Ge\kern -0.1em V\!/}c^2}\xspace}

\def\barn{\aunit{b}\xspace}

\def\fb   {\ensuremath{\aunit{fb}}\xspace}
\def\invfb   {\ensuremath{\fb^{-1}}\xspace}

\def\gsim{{~\raise.15em\hbox{$>$}\kern-.85em
          \lower.35em\hbox{$\sim$}~}\xspace}
\def\lsim{{~\raise.15em\hbox{$<$}\kern-.85em
          \lower.35em\hbox{$\sim$}~}\xspace}

\def\tell1  {TELL1\xspace}
\def\ukl1   {UKL1\xspace}

\newcommand{\lhcborcid}[1]{\href{https://orcid.org/#1}{\hspace*{0.1em}\raisebox{-0.45ex}{\includegraphics[width=1em]{figs/orcidIcon.pdf}}}}

\def\DorDsp  {{\ensuremath{\D^+_{(s)}}}\xspace}

\def\DTohll     {\decay{\DorDsp}{\Ph^{\pm}\ell^{+}\ell^{(\prime)\mp}}}

\def\DspToKmumuSS   {\decay{\Dsp}{\Km\mup\mup}}

\def\be{\begin{equation}}
\def\ee{\end{equation}}
\def\beq{\begin{eqnarray}}
\def\eeq{\end{eqnarray}}

\newcommand\dmumu{\decay{\Dz}{\mup\mun}}

\def\brlimitninety{\ensuremath{3.1 \times 10^{-9}}\xspace} 

\newcommand{\Ksmumu}{\ensuremath{\Ks\to\mu^+\mu^-}\xspace}
\newcommand{\Ks}{\KS}

\newcommand{\KzFM}{\ensuremath{K^0_{\mathrm{S(L)}}\to\mu^+\mu^-\mu^+\mu^-}\xspace}
\newcommand{\BRof}[1]{\ensuremath{{\cal B}(#1)}\xspace}

\newcommand{\KFM}{\ensuremath{\Ks\to\mu^+\mu^-\mu^+\mu^-}\xspace}
\newcommand{\KlFM}{\ensuremath{\Kl\to\mu^+\mu^-\mu^+\mu^-}\xspace}

\newcommand{\KLFM}{\ensuremath{\Kl\to\mu^+\mu^-\mu^+\mu^-}\xspace}
\newcommand{\Kl}{\KL}

\def\sigmapmumu{\ensuremath{\PSigma^+ \to \Pp \mu^+ \mu^-}\xspace}

\def\B{\ensuremath{\mathcal{B}}\xspace}

\def\signdefault{\ensuremath{4.1}\xspace}
\def\brmeasured{\ensuremath{(2.2\,^{+\,1.8}_{-\,1.3})\times 10^{-8}}\xspace}

\def\LcTopmumu  {\decay{\Lc}{\proton \mumu}}

\newcommand{\Dkkmm}{\mbox{\ensuremath{\Dz\to\Kp\Km\mu^+\mu^-}}\xspace}
\newcommand{\Dppmm}{\mbox{\ensuremath{\Dz\to\pip\pim\mu^+\mu^-}}\xspace}

\usepackage{lineno}
\numberwithin{equation}{section}
\definecolor{email}{rgb}{0.00,0.00,0.84}
\begin{document}
\setcounter{page}{1}
\date{} 
\date{March 28, 2024}
\title{\large \bf 12th Workshop on the CKM Unitarity Triangle\\ Santiago de Compostela, 18-22 September 2023 \\ \vspace{0.3cm}
\LARGE Rare Strange and Charm Decays \\at LHCb}
\author[]{Paras Naik \\ \href{mailto:paras.naik@cern.ch}{\color{email} paras.naik@cern.ch}}
\affil[]{Oliver Lodge Laboratory, University of Liverpool \\ Liverpool,
United Kingdom \\ ~ \\ On behalf of the LHCb collaboration}
\maketitle
\begin{abstract} 
 Thanks to the large cross-section for the production of strange and charm mesons at the Large Hadron Collider, LHCb has recently performed several new explorations of rare transitions that are highly suppressed or forbidden by the Standard Model: mostly flavour changing neutral current decays, but also lepton flavour violating decays.
Physics beyond the Standard Model could contribute at tree and loop levels.
\end{abstract} \maketitle

\section{Rare strange decays at LHCb}

\subsection{\texorpdfstring{\boldmath $K^0_S\to \mu^+ \mu^-$}{}}

\noindent The decay \Ksmumu is a flavor-changing neutral current (FCNC) process which has not been observed yet.
In the Standard Model (SM), this decay is highly suppressed~\cite{Ecker:1991ru,Isidori:2003ts}, with an expected branching fraction~\cite{DAmbrosio:2017klp} of: 
\begin{equation}
  \BRof\Ksmumu_{\rm SM} = (5.18 \pm 1.50_{\rm LD} \pm 0.02_{\rm SD}) \times 10^{-12},
\end{equation}
where LD(SD) uncertainties relate to long-distance(short-distance) effects.

A search for the rare decay \Ksmumu has  been  performed  on  a LHCb data set of about $8.6\invfb$~\cite{LHCb:2020ycd}.
The data are consistent both with the background-only hypothesis and the combined background and SM signal expectation at the  $1.4 \sigma$ and $1.3\sigma$ level, respectively.
We set the most stringent upper limit to date on the \Ksmumu branching fraction --- $2.1\,(2.4)\times 10^{-10}$ at 90\,(95)\% confidence level, improving the previous best limit by a factor of four. The obtained results supersede those  of  our  previous publications~\cite{LHCb:2017qna,LHCb:2012vqi}.

We expect to improve limits on $\mathcal{B}(K^0_{S}\to  \mu^+ \mu^-)$  by an order of magnitude after LHCb Upgrade I (50 fb$^{-1}$)~\cite{LHCb:2018roe}. After LHCb Upgrade II (300 fb$^{-1}$) we expect to reach a sensitivity close to the SM expectation~\cite{LHCb:2018roe}.

\subsection{\texorpdfstring{\boldmath $K^0_{S(L)}\to \mu^+ \mu^- \mu^+ \mu^-$}{}}

\noindent The $K^0_{S(L)}\to \mu^+ \mu^- \mu^+ \mu^-$ decays are flavour-changing neutral current processes that have not yet been observed. In the SM, their decay rates are highly suppressed, with expected branching fractions~\cite{DAmbrosio:2013qmd}:
\begin{equation}
  \BRof\KFM_{\text{SM}} \sim (1 \text{--} 4) \times 10^{-14},
\end{equation}
\begin{equation}
  \BRof\KLFM_{\text{SM}} \sim (4 \text{--} 9)  \times 10^{-13}.
\end{equation}
\noindent Since the expected LHCb acceptance for \Kl decays is $\sim 2\times10^{-3}$ times smaller than the \Ks one~\cite{Junior:2018odx},
the SM \KlFM yield expected in the experiment is small compared to that of the \Ks decay. 

Physics beyond the SM can lead to large enhancements of \BRof\KFM with respect to the SM prediction. For instance, proposed dark-sector scenarios can enhance the branching fractions 
up to 
around $2\times10^{-12}$~\cite{Goudzovski:2022vbt,Hostert:2020xku}. 

A search for \KzFM has been performed analysing $5.1\invfb$ of LHCb data recorded from 2016 to 2018~\cite{LHCb:2022tpr}.  No signal is observed.
The upper limits at 90\% C.L. are found to be:
\begin{align}
    \BRof\KFM &< 5.1 \times 10^{-12}, \nonumber \\
    \BRof\KlFM &< 2.3 \times 10^{-9} \nonumber .
\end{align}
The obtained upper limits are the first reported for the \KzFM decay modes. The observed values are close to the maximum values allowed in the dark photon models~\cite{Goudzovski:2022vbt}.
LHCb could achieve sensitivities at the level of the SM prediction for \BRof\KFM using 300\invfb of integrated luminosity (available after LHCb Upgrade II)~\cite{LHCb:2022tpr}.

\subsection{\texorpdfstring{\boldmath \sigmapmumu}{}}

A search for the \sigmapmumu decay was performed by  
LHCb using $3 \invfb$ of $pp$ collisions at centre-of-mass energies $\sqrt{s} = 7$ and $8 \tev$ (Run 1)~\cite{LHCb:2017rdd}. Evidence for the 
\sigmapmumu decay was found with a significance of $\signdefault$ standard deviations, 
including systematic uncertainties.
A branching fraction $\mathcal{B}(\sigmapmumu) = \brmeasured$ was measured, 
consistent with the SM prediction. An LHCb Run 2 update of this analysis is in progress.

\subsection{Future analyses}

\noindent Besides the analyses discussed above, the following interesting channels are expected to be explored by LHCb in the future:
\begin{itemize}
\item $K^0_{S}\to  \pi^+ \pi^- \mu^+ \mu^-$ --- This decay is highly constrained by phase space.
\item $K^0_{S}\to (\pi/\mu/e)^+ (\pi/\mu/e)^- e^+e^-$ --- These are interesting channels, requiring further data due to low electron efficiencies.
\item $K^0_{S}\to e\mu$, $K^+ \to \pi^+e\mu$ --- These channels allow the study of potential lepton flavor violation.
\end{itemize}

\section{Rare charm decays at LHCb}

\subsection{\texorpdfstring{\boldmath Search for rare and forbidden $D^+_{(s)}$  decays}{}}

\noindent Decays of the form \DTohll, where $\Ph$ is a kaon or pion and $\ell^{(\prime)}$ is an electron or muon, are rare or forbidden within the SM, making room for new physics enhancements.
Searches have been made for a total of 25 
decays 
using \SI{1.6}{\per\femto\barn} of \proton\proton collision data collected by the LHCb experiment during 2016~\cite{LHCb:2020car}.
No significant deviations from the background-only hypotheses are seen.
The \SI{90}{\percent} CL limits on the branching fractions vary from \num{1.4e-8} to \num{6.4e-6}
and represent the world's best limits for 23 of these decays. In the majority of the channels the improvement on previous limits is more than an order of magnitude, with the largest improvement being a factor of five hundred in \DspToKmumuSS decays.

\subsection{\texorpdfstring{\boldmath Search for $D^0 \to \mu^+ \mu^-$ decays}{}}

\noindent A search for the \dmumu decay in data corresponding to $9\invfb$ of $pp$ collision data collected by the LHCb experiment is performed~\cite{LHCb:2022jaa}. No excess with respect to the background expectation has been found and an upper limit of 
$\mathcal{B}(\dmumu) < \brlimitninety$ at 90\% CL has been set. 
This result represents an improvement of more than a factor two with respect to the previous LHCb result. This measurement constitutes the most stringent limit on the relevant FCNC couplings in the charm sector, allowing additional constraints on physics models beyond the SM that predict the branching fraction of \dmumu. 

By the end of LHCb Run 3 
(23 fb$^{-1}$) [LHCb Upgrade II (300 fb$^{-1}$)], we expect to achieve a sensitivity of $\mathcal{B}(D^0 \to \mu^+ \mu^-) < 
5.9 \times 10^{-10}~[
1.8 \times 10^{-10}$]~\cite{LHCb:2018roe}. 

\subsection{\texorpdfstring{\boldmath $D^0 \to h^+ h^- \mu^+ \mu^-$  angular analysis}{}}

\noindent A measurement of the full set of \CP-averaged angular observables and their \CP asymmetries in \Dppmm and \Dkkmm decays is performed~\cite{LHCb:2021yxk}. The analysis uses \proton\proton collision data collected with the \lhcb detector at center-of-mass energies of 7, 8, and 13\tev, corresponding to an integrated luminosity of 9\invfb. This is the first full angular analysis of a rare charm decay ever performed. The full set of \CP-averaged angular observables and their asymmetries are measured as a function of the dimuon invariant mass. The results are consistent with expectations from the SM and with \CP
 symmetry. The measured null-test observables 
 are in agreement with the SM null hypothesis with overall $p$ values of 79\% (0.8\%) for \Dppmm (\Dkkmm) decays, corresponding to 0.3 (2.7) Gaussian standard deviations. These measurements will help to constrain the parameter space of physics models extending the SM. 

\subsection{\texorpdfstring{\boldmath Search for $D^{*0} \to \mu^+ \mu^-$ decays}{}}

\begin{figure}[!tb]
\centering
\includegraphics[width=0.50\textwidth]{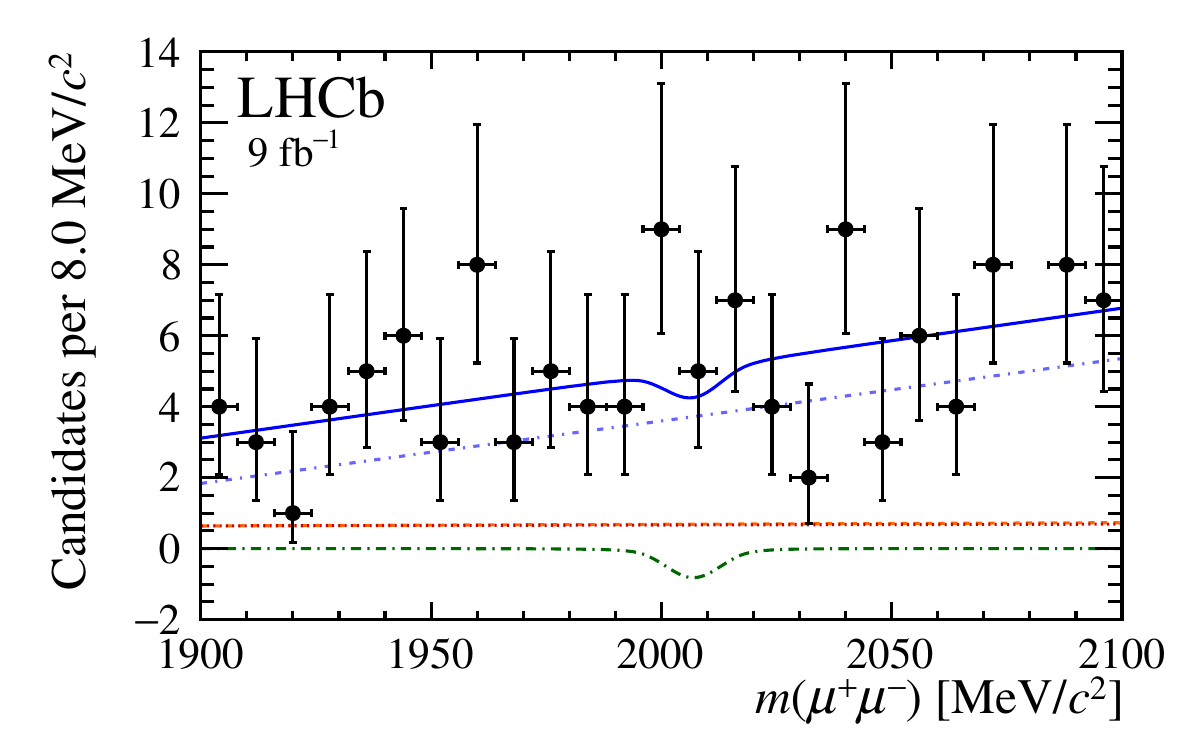}\hfill
\includegraphics[width=0.50\textwidth]{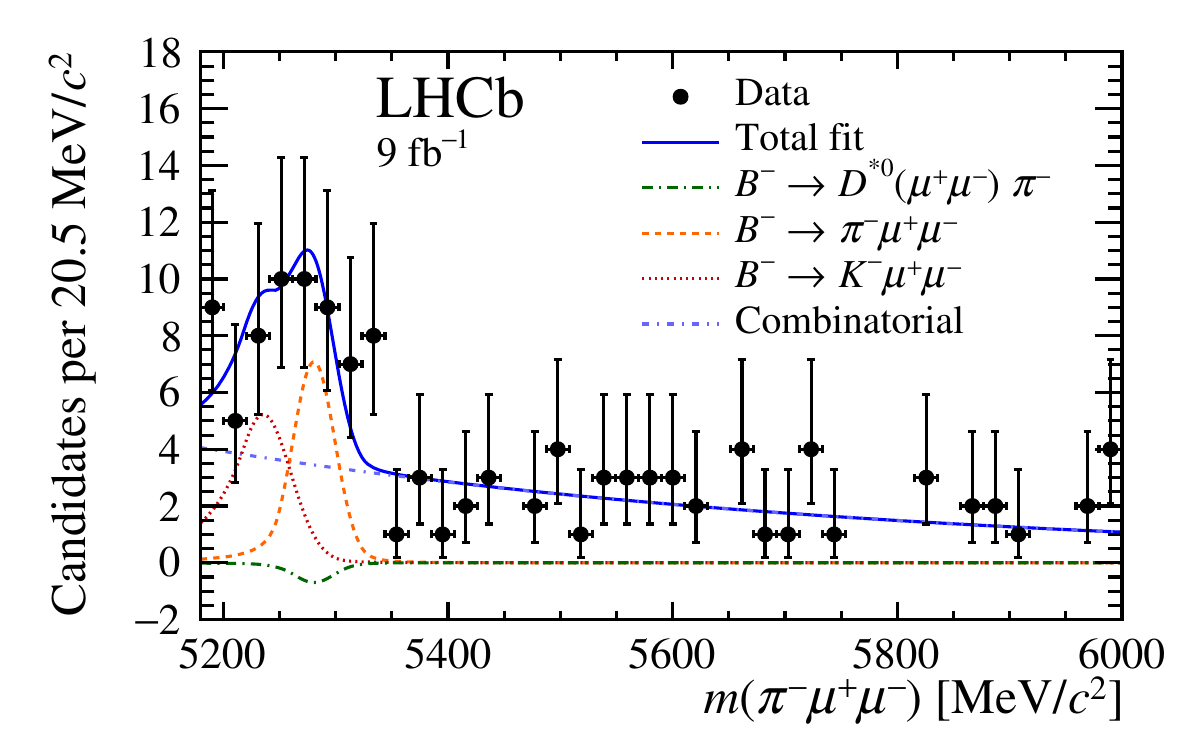}
\caption{
    Reconstructed (left) $\mumu$ and (right) $\pim\mumu$ invariant-mass distributions for the selected \mbox{\decay{B^-}{\Dstarz(\mumu)\,\pim}} candidates, with fit results superimposed. 
}
\label{fig:mass:dstar} 
\end{figure}

A search for the \decay{\Dstarz}{\mumu} decay is performed by analysing \decay{B^-}{\pim\mumu} decays~\cite{LHCb:2023fuw}.  
The analysis uses a data sample corresponding to an integrated luminosity of $9\invfb$ collected with the LHCb experiment in $pp$ collisions between 2011 and 2018. This is the first search for a rare charm-meson decay exploiting its production in beauty-meson decays. 
Figure~\ref{fig:mass:dstar} shows the dimuon and $B$-candidate invariant-mass distributions of selected \decay{B^-}{\Dstarz\pim} candidates, with results of the fit superimposed.
No excess with respect to the
background-only hypothesis is observed and an upper limit of $\BF(\decay{\Dstarz}{\mumu}) < 2.6 \times 10^{-8}$ at $90\%$~CL is set. 
This measurement is the first limit on the branching fraction of the \decay{\Dstarz}{\mumu} decay and the most stringent limit on \Dstarz decays to leptonic final states.

\subsection{\texorpdfstring{\boldmath \LcTopmumu}{}}
\noindent  A search for the \LcTopmumu decay was performed, using 3 \invfb of $pp$ data collected by LHCb~\cite{LHCb:2017yqf}. The analysis is performed in three regions of dimuon mass: $\phi$, $\omega$ and nonresonant. The upper limit on the nonresonant mode, $\BRof\LcTopmumu < 7.7~(9.6)\times 10^{-8}~{\rm at}~90(95)\%$ confidence level, is improved by two orders of magnitude with respect to the previous measurement~\cite{BABAR}. For the first time the signal is observed in the $\omega$ region, with a statistical significance of 5 standard deviations. 
An LHCb Run 2 update of this analysis in progress.

\section{Summary}

LHCb has worked ``like a charm'' but seen ``nothing strange'' yet. LHCb in its first two runs has published several rare strange and charm results, now often providing the best available limits. LHCb is continuing to exploit its unique data set to investigate rare decays. Expect more LHCb rare decay results in the LHCb Upgrade eras.



\bibliographystyle{amsplain}

\end{document}